\documentclass[showpacs,amsmath,amssymb]{article}
\usepackage{amsfonts}
\usepackage{bm}

\newcommand{{\Cd}}{{\mathbb{C}^d}}

\def\<{\langle}
\def\>{\rangle}

\begin{document}
\title{\textbf{Memory in a nonlocally damped oscillator}}
\author{Dariusz Chru\'sci\'nski and
Jacek Jurkowski\footnote{Corresponding author. E-mail address: jacekj@fizyka.umk.pl}\\[1ex]
Institute of Physics, Nicolaus Copernicus University,\\
Grudzi\c{a}dzka 5/7, 87--100 Toru\'n, Poland}
\date{}

\maketitle

\begin{abstract}

We analyze the new equation of motion for the damped oscillator.
It differs from the standard one by a damping term which is
nonlocal in time and hence it gives rise to a system with memory.
Both classical and quantum analysis is performed. The
characteristic feature of this nonlocal system is that it breaks
local composition low for the classical Hamiltonian
dynamics and the corresponding quantum propagator.\\[1ex]
\end{abstract}

\section{Introduction}

The damped harmonic oscillator is one of the simplest quantum
systems displaying the dissipation of energy. Moreover, it is of
great physical importance and has found many applications
especially in quantum optics. For example it plays a central role
in the quantum theory of lasers and masers \cite{QO1,QO2,QO3}. The
standard Hilbert space formulation of Quantum Mechanics gives rise
to the unitary evolution with no room for dissipative behavior.
The usual approach to include dissipation is the quantum theory of
open systems \cite{Haake,Davies,Petruccione,Carmichael,Tarasov}.
In this approach the dynamics of a quantum system is no longer
unitary but it is defined by a semigroup of completely positive
maps in the space of density operators \cite{Alicki}.

Another strategy to describe dissipative quantum systems is based
on the old idea of Bateman \cite{Bat31}. Bateman has shown that in order to
apply the standard canonical formalism of classical mechanics to
dissipative and non-Hamiltonian systems, one can double the number
of degrees of freedom, so as to deal with an effective isolated
classical Hamiltonian system. In this approach the new degrees of
freedom may be interpreted as variables describing a reservoir.
Applying this idea to a damped harmonic oscillator one obtains a
pair of damped oscillators (so called Bateman's dual system): a
primary one and its time reversed image. The Bateman dual
Hamiltonian has been rediscovered by Morse and Feshbach
\cite{Mor53} and Bopp \cite{Bopp73} and the detailed quantum
mechanical analysis was performed by Feshbach and Tikochinski
\cite{FT}. The quantum Bateman system was then analyzed by many
authors (see for the historical review \cite{Dekker81}). Recently
it was investigated  in connection with quantum field theory and
quantum groups \cite{Vit92} (see also \cite{Vit94,Vit95}).
Finally, the detailed analysis of spectral properties of the
Bateman system was recently performed in \cite{Ann1,Ann2} (see
also \cite{Bla04}). For other approaches to quantization of damped
oscillator see also recent papers \cite{MM,Dito,L}.

In the present paper we propose a different approach. Instead of
analyzing the standard equation of motion for the damped
oscillator
\begin{equation}\label{damped}
 \ddot{x}+2\gamma\dot{x}+\omega^2 x=0\ ,
\end{equation}
where $\gamma$ is a positive damping constant, and $\omega$ is a
frequency of the undamped motion, we shall study  a modified
equation
\begin{equation}\label{damped_NEW}
 \ddot{x}+2\gamma\tanh(\gamma(t-t_0))\dot{x}+\omega^2 x=0\ ,
\end{equation}
where $t_0$ is an arbitrary constant. It is clear that in the
 limit `$\gamma t \rightarrow \infty$',
formula (\ref{damped_NEW}) reproduces the standard equation
(\ref{damped}). Therefore, both dynamics may differ for short
times only, that is, for `$t$' satisfying
\[t \ll 1/\gamma\ . \]

What is the origin of (\ref{damped_NEW})? It turns out that both
(\ref{damped}) and (\ref{damped_NEW}) give the same frequency of
oscillations of the damped motion\footnote{Throughout the paper we
assume $\omega > \gamma$.}
\begin{equation}\label{Omega}
\Omega=\sqrt{\omega^2-\gamma^2}\ .
\end{equation}
Moreover, there is no other linear equation of motion with this
property. Interestingly, equation (\ref{damped_NEW}) introduces
additional parameter $t_0$. It turns out that $t_0$ is responsible
for an effective memory, that is, in addition to a `local time
$t$' evolution governed by (\ref{damped_NEW}) `remembers' about
its history at $t_0$. It means that the initial value problem for
(\ref{damped_NEW}) has to be supplemented by initial data
$x(t_0)=x_0$ and $\dot{x}(t_0) = v_0$. There is a crucial
difference between (\ref{damped}) and (\ref{damped_NEW}): the
initial data for the standard `local' equation may be provided for
arbitrary initial moment of time. It is no longer true for the
modified `nonlocal' equation: now the initial moment necessarily
equals $t_0$. Changing initial moment $t_0$ one changes the
corresponding equation (\ref{damped_NEW}) and hence the whole
evolution of our damped system. This property is responsible for a
peculiar feature of (\ref{damped_NEW}). The composition law for
classical dynamics is no longer true, that is, evolution from
$t_0$ to $t_1$, and then from $t_1$ to $t_2$ differs from the
direct evolution from $t_0$ to $t_2$.

Now, the quantum system corresponding to (\ref{damped_NEW}) may be
easily constructed. Our damped system turns out to be governed by
nonlocal Hamiltonian and hence the standard composition law for
quantum propagators
\begin{equation}\label{KKK}
    K(x_2,t_2;x_0,t_0) = \int dx_1 K(x_2,t_2;x_1,t_1)
    K(x_1,t_1;x_0,t_0) \ ,
\end{equation}
does not hold.  We stress that the local composition law
(\ref{KKK}) is satisfied by the quantum propagator derived from
the standard equation (\ref{damped}) via well known
Caldirola-Kanai approach \cite{Caldirola,Kanai}. The violation of
(\ref{KKK}) defines the basic difference between two descriptions
based on (\ref{damped}) and (\ref{damped_NEW}), respectively.  We
hope that our analysis sheds new light on this old problem.

\section{Hamiltonian formulation}

There were many attempts for hamiltonian formulation of Newton
equation for systems with one degree of freedom (see e.g.
\cite{Kanai,Hav57,Kob79,Lop96,Lop05}).
 It is well known that (\ref{damped}) may
be derived from the so called Caldirola-Kanai Hamiltonian
\cite{Caldirola,Kanai}
\begin{equation}\label{H_Kanai}
H(x,p,t)\;=\;\frac12 m_0\omega^2x^2\,e^{2\gamma t}+\frac{1}{2m_0}
p^2\,e^{-2\gamma t}\ .
\end{equation}
It is clear that Caldirola-Kanai Hamiltonian may be considered as
the standard Hamiltonian of the harmonic oscillator with
time-dependent  mass given by
\begin{equation} \label{m-CK}
 m(t)=m_0 e^{2\gamma t}\ .
\end{equation}
Consider now a family of Hamiltonians with an arbitrary
time-dependent mass $m(t)$:
\begin{equation} \label{H_m}
 H(x,p,t)=\frac{p^2}{2m(t)}+\frac12 m(t)\omega^2 x^2\ .
\end{equation}
The corresponding  Newton equation reads as follows
\begin{equation}\label{N_m}
 \ddot{x}+\frac{\dot{m}(t)}{m(t)}\,\dot{x}+\omega^2 x=0\ .
\end{equation}
Let us assume that in the asymptotic regime `$\gamma t \rightarrow
\infty$', equation (\ref{N_m}) reproduces the standard equation
(\ref{damped}), that is
\begin{equation}\label{cond}
\kappa(t) := \frac 12\, \frac{\dot{m}(t)}{m(t)}\
{\longrightarrow}\ \gamma\ .
\end{equation}
To solve (\ref{N_m}) one represents complexified $x(t)$ as follows
\begin{equation}\label{}
x(t)=\lambda(t)\,e^{i\phi(t)}\ ,
\end{equation}
and obtains the following equations for time-dependent real
quantities $\lambda,\phi$ and $\kappa$:
\begin{eqnarray}\label{eq11}
 \ddot{\lambda}-\lambda\dot{\phi}^2+2\kappa\dot{\lambda}+\omega^2\lambda &=& 0\ , \\
 \ddot{\phi}\lambda+2\dot{\phi}\dot{\lambda}+2\kappa\dot{\phi}\lambda &=&
 0\, ,\label{eq12}
\end{eqnarray}
where we omitted  explicit time-dependence. Now, we make our basic
assumption that $\dot{\phi} = \Omega$ with $\Omega$ defined in
(\ref{Omega}). Equation (\ref{eq12}) reduces then to
\[
 \dot{\lambda}+\kappa\lambda=0\ ,
\]
and hence equation (\ref{eq11}) implies
\[ \Omega^2 =
\omega^2-\kappa^2-\dot{\kappa}\ ,
\]
which together with (\ref{Omega}) gives finally
\[
 \dot{\kappa}+\kappa^2=\gamma^2\ .
\]
The above equation has the following solutions
\begin{equation}
 \kappa(t) \;=\;\gamma \tanh(\gamma(t-t_0))\quad\mbox{ or }\quad
\kappa(t)\;=\;\pm\gamma\ ,
\end{equation}
with $t_0$ being an integration constant. Note, however, that due
to (\ref{cond}) the solution $\kappa(t)=-\gamma$ is not
acceptable. It is clear that $\kappa(t)=\gamma$ corresponds to the
Caldirola-Kanai model with $m(t)=e^{2\gamma t}$, whereas
$\kappa(t)= \gamma \tanh(\gamma(t-t_0))$ gives rise to our new
equation (\ref{damped_NEW}) described by a Hamiltonian  with
time-dependent mass
\begin{equation} \label{m}
 m(t)=m_0\cosh^2(\gamma(t-t_0))\ ,
\end{equation}
satisfying $m(t_0)=m_0$. We may therefore interpret
(\ref{damped_NEW}) as a Newton equation for a particle with
constant mass $m_0$ under the action of the time dependent
nonlocal force
\[   F(t;t_0) = - 2m_0\gamma\tanh(\gamma(t-t_0))\dot{x}- m_0\omega^2 x\ ,\]
or as a Newton equation for a particle with time-dependent
nonlocal mass (\ref{m}) under the action of  perfectly local force
$-m_0\omega^2 x$.

\section{Classical dynamics}

Consider now classical Hamiltonian dynamics generated by
(\ref{N_m}) with $m(t)$ defined in (\ref{m}). It is given by
\begin{equation}\label{C-nasz}
   \left( \begin{array}{c} x(t) \\ p(t) \end{array} \right) =
   \Lambda^{\rm new}(t;t_0) \left( \begin{array}{c} x_0 \\ p_0 \end{array}
   \right) \ ,
\end{equation}
where the $2 \times 2$ matrix
\[ \Lambda^{\rm new}(t;t_0) \ = \ \left( \begin{array}{cc} \Lambda^{\rm new}_{xx} &
\Lambda^{\rm new}_{xp} \\
\Lambda^{\rm new}_{px} & \Lambda^{\rm new}_{pp} \end{array}
   \right) \ ,  \]
reads as follows
\begin{equation}\label{}
\Lambda^{\rm new}_{xx} =
\frac{\cos(\Omega(t-t_0))}{\cosh(\gamma(t-t_0))} \ , \qquad
\Lambda^{\rm new}_{xp} = \frac{1}{m_0\Omega}
\frac{\sin(\Omega(t-t_0))}{\cosh(\gamma(t-t_0))} \ ,
\end{equation}
and
\begin{eqnarray}\label{}
\Lambda^{\rm new}_{px} &=& - m_0 \Big[
\Omega\cosh(\gamma(t-t_0))\sin(\Omega(t-t_0)) +
\gamma\sinh(\gamma(t-t_0))\cos(\Omega(t-t_0)) \Big] \ , \\
  \Lambda^{\rm new}_{pp} &=&
\Omega\cosh(\gamma(t-t_0))\cos(\Omega(t-t_0)) -
\frac{\gamma}{\Omega} \sinh(\gamma(t-t_0))\sin(\Omega(t-t_0)) \ .
\end{eqnarray}
The crucial feature of (\ref{C-nasz}) is that it breaks the  local
composition law
\begin{equation}\label{LLL}
    \Lambda(t_2;t_0) = \Lambda(t_2;t_1) \circ \Lambda(t_1;t_0) \
    ,
\end{equation}
for $t_0 \leq t_1 \leq t_2$, that is, formula (\ref{LLL}) does not
hold for $\Lambda=\Lambda^{\rm new}$.

Let us compare our new evolution (\ref{C-nasz}) with the classical
dynamics of the Caldirola-Kanai system. Note, that we may
trivially change the formula (\ref{m-CK}) for time-dependent mass
\begin{equation}\label{m-new}
    m(t) = m_0 e^{2\gamma(t-t_0)}\ ,
\end{equation}
without affecting the equation of motion. Now, these two models
contain additional parameter $t_0$. One easily finds for the
corresponding Caldirola-Kanai dynamics $\Lambda^{\rm CK}(t;t_0)$:
\begin{eqnarray} \label{La}
\Lambda^{\rm CK}_{xx} &=& \Big[\cos(\Omega(t-t_0))+\frac{\gamma}{\Omega}\sin(\Omega(t-t_0))\Big]e^{-\gamma(t-t_0)},\\
\Lambda^{\rm CK}_{xp} &=& \frac{1}{m_0\Omega}\sin(\Omega(t-t_0))e^{-\gamma(t-t_0)},\\
\Lambda^{\rm CK}_{px} &=& -m_0\Omega\Big(1+\frac{\gamma^2}{\Omega^2}\Big)\sin(\Omega(t-t_0))e^{\gamma(t-t_0)},\\
\label{Lc}\Lambda^{\rm CK}_{pp} &=&
\Big[\cos(\Omega(t-t_0))-\frac{\gamma}{\Omega}\sin(\Omega(t-t_0))\Big]
e^{\gamma(t-t_0)}.
\end{eqnarray}
One may check that $\Lambda^{\rm CK}$ satisfies
\begin{equation}\label{LLL-local}
    \Lambda^{\rm CK}(t_2;t_0) = \Lambda^{\rm CK}(t_2;t_1) \circ \Lambda^{\rm CK}(t_1;t_0) \
    ,
\end{equation}
for $t_0 \leq t_1 \leq t_2$. Note, that computing $\Lambda^{\rm
CK}(t_2,t_1)$ one applies (\ref{La})--(\ref{Lc}) with $t_0$
replaced by $t_1$, and $m_0$ replaced by $m_1 := m(t_1) =
m_0e^{2\gamma(t_1-t_0)}$. Then formula (\ref{LLL-local}) easily
follows as a simple consequence of $e^{a+b}=e^ae^b$. It is no
longer true if we replace (\ref{m-new}) by (\ref{m}).

Finally,  let us compare both approaches in the simplified
situation corresponding to pure dissipation, that is, $\omega =0$.
In this case the corresponding formula for $\Lambda^{\rm
new}(t;t_0)$ considerably simplifies
\begin{equation}\label{nasz-pure}
    \Lambda^{\rm new}_0(t;t_0) \ = \ \left( \begin{array}{cc} 1 &
(m_0\gamma)^{-1}\tanh(\gamma(t-t_0)) \\
0 & 1 \end{array}
   \right) \ .
\end{equation}
However, $\Lambda^{\rm new}_0$ still breaks  (\ref{LLL}). In the
case of Caldirola-Kanai dynamics one finds
\begin{equation}\label{CK-pure}
    \Lambda^{\rm CK}_0(t;t_0) \ = \ \left( \begin{array}{cc} 1 &
 (m_0\gamma)^{-1}\sinh(\gamma(t-t_0))\, e^{-\gamma(t-t_0)} \\ 0 & 1
\end{array}
   \right) \ .
\end{equation}
Note, that in a asymptotic regime $\gamma t \longrightarrow
\infty$ one obtains
\begin{equation}\label{nasz-pure}
    \Lambda^{\rm new}_{\rm asympt} \ = \ \left( \begin{array}{cc} 1 &
(m_0\gamma)^{-1} \\
0 & 1 \end{array}
   \right) \  , \ \ \ \ \  \Lambda^{\rm CK}_{\rm asympt} \ = \ \left( \begin{array}{cc} 1 &
 (2m_0\gamma)^{-1} \\ 0 & 1
\end{array}
   \right) \ ,
\end{equation}
that is, both dynamics have different asymptotic states: $(x_0 +
p_0/(m_0\gamma),p_0)$ and $(x_0 + p_0/(2m_0\gamma),p_0)$ for the
Caldirola-Kanai system.

\section{Quantum propagator}

Consider now a quantum propagator corresponding to Hamiltonian
(\ref{H_m}) with $m(t)$ given by (\ref{m}). Since the Hamiltonian
is quadratic in $(x,p)$ one easily finds for the propagator (see
e.g. \cite{Kor})
\begin{equation}\label{PROPAGATOR}
K(x,t;x_0,t_0) = N(t-t_0)\, A(x,t;x_0,t_0)\cdot B(x,t;t_0) \ ,
\end{equation}
where
\begin{eqnarray}\label{AB}
     A(x,t;x_0,t_0) & = & \exp\left\{\frac{i}{2\hbar}\frac{m_0\Omega}{\sin(\Omega(t-t_0))}
\Big[(x_0^2+x^2)\cos(\Omega(t-t_0)) -2x_0x\cosh(\gamma(t-t_0))\Big]\right\} \ ,\nonumber\\
  B(x,t;t_0) & = &
  \exp\left\{\frac{i}{2\hbar}\, m_0x^2\,
  \frac{\sinh(\gamma(t-t_0))}{\sin(\Omega(t-t_0))} \Big[
  \Omega\cos(\Omega(t-t_0))\sinh(\gamma(t-t_0))\right.\nonumber\\
& & \left.\hspace*{10mm}
-\;\gamma \sin(\Omega(t-t_0))
  \cosh(\gamma(t-t_0)) \Big]\right\} \ ,
\end{eqnarray}
and the normalization factor is given by
\begin{equation}
N(t-t_0)=\sqrt{\frac{m_0\Omega\cosh(\gamma(t-t_0))}{2\pi
i\hbar\sin(\Omega(t-t_0))}} \ .
\end{equation}
 Note, that when $\gamma=0$ the factor $B$ equals to 1 and one recovers standard propagator
for the harmonic oscillator \cite{Feynman}. On the other hand when
$\omega=0$, that is, in a pure damping case, formula
(\ref{PROPAGATOR}) reduces to
\begin{eqnarray}\label{K-gamma}
     K_0(x,t;x_0,t_0) = N_0(t-t_0)\exp\Big\{\frac{i\gamma}{2\hbar}\frac{m_0(x-x_0)^2}{\tanh(\gamma(t-t_0))}
     \Big\}\ ,
\end{eqnarray}
with
\begin{equation}
N_0(t-t_0)= \sqrt{\frac{\gamma m_0}{2\pi i
\hbar\tanh(\gamma(t-t_0))}} \ .
\end{equation}
For $\gamma=0$ formula (\ref{K-gamma}) reproduces quantum
propagator for a free particle \cite{Feynman}.
Interestingly, the formula (\ref{K-gamma}) for a quantum
propagator for a particle undergoing linear damping was recently
derived by Kochan \cite{Kochan} by using purely geometrical
considerations. Here we showed that it may be derived via standard
methods by taking suitable nonlocal Hamiltonian with memory. In
\cite{Kochan} the formula for a full propagator of the damped
oscillator was also presented which however differs from
(\ref{PROPAGATOR}). Actually, Kochan propagator is given by
\begin{equation}\label{K-Kochan}
K_{\rm Kochan}(x,t;x_0,t_0) = N(t-t_0)\, A(x,t;x_0,t_0) \ ,
\end{equation}
and therefore differs from (\ref{PROPAGATOR}) only by a factor
$B(x,t;t_0)$. Both propagators $K$ and $K_{\rm Kochan}$ produce
the same limits for $\gamma=0$ (harmonic oscillator) and for
$(\gamma=0\ \&\ \omega=0)$ (free particle). Observe that
$B(x,t;t_0)$ produces only an additional phase factor at the final
position `$x$'. Therefore both formulae give the same results for
a mean position of a particle: starting from the initial wave
function $\psi(x_0,t_0)$ one finds
\begin{equation}\label{}
\psi(x,t;t_0)=\int\limits_{-\infty}^\infty
K(x,t;x_0,t_0)\,\psi(x_0,t_0)dx_0\ ,
\end{equation}
and
\begin{eqnarray}\label{}
\psi_{\rm Kochan}(x,t;t_0) &=& \int\limits_{-\infty}^\infty K_{\rm
Kochan}(x,t;x_0,t_0)\,\psi(x_0,t_0)dx_0 \;=\;
\frac{1}{B(x,t;t_0)}\, \psi(x,t;t_0)\ ,
\end{eqnarray}
and hence
\begin{eqnarray}\label{}
    \< x(t;t_0) \> &=& \int\limits_{-\infty}^\infty
x\,|\psi(x,t;t_0)|^2dx \;=\; \int\limits_{-\infty}^\infty
x\,|\psi_{\rm Kochan}(x,t;t_0)|^2dx\ .
\end{eqnarray}
It is easy to show that $\< x(t;t_0) \>$ satisfies our modified
nonlocal equation (\ref{damped_NEW}):
\begin{equation}\label{}
     \frac{d^2}{dt^2}\< x(t;t_0) \>\, +\, 2\gamma\tanh(\gamma(t-t_0))\frac{d}{dt}\< x(t;t_0) \>
     \, +\,\omega^2 \< x(t;t_0) \>=0\ .
\end{equation}
In the Appendix we show that this additional phase factor
$B(x,t;t_0)$ may be eliminated by a suitable canonical
transformation. The price we pay for a simpler formula for a
propagator is the much more complex form of the Hamiltonian which
apart from the standard quadratic terms $x^2$ and $p^2$ contains
also a mixed term $xp$ (see Appendix).

Let us compare the above formulae with the propagator for the
Caldirola-Kanai system. One easily finds the CK propagator
corresponding to $m(t)$ given by (\ref{m-new})
\begin{eqnarray}\label{P_KH}
     K_{\rm CK}(x,t;x_0,t_0) &=& N_{\rm CK}(t-t_0)\exp\Big\{
\frac{i}{2\hbar}\,m_0\Big[\gamma\Big(x_0^2e^{2\gamma
t_0}-x^2e^{2\gamma t}\Big)  \nonumber\\
  &+&    \Omega\cot(\Omega(t-t_0))\Big(x_0^2e^{2\gamma t_0}
+  x^2e^{2\gamma t}\Big)
-\;2\frac{\Omega}{\sin(\Omega(t-t_0))}xx_0e^{\gamma(t+t_0)}\Big]\Big\},
\end{eqnarray}
where the normalization factor
\begin{equation}
N_{\rm CK}(t-t_0)= \sqrt{\frac{m_0\Omega\,e^{\gamma(t-t_0)}}{2\pi
i\hbar\sin(\Omega(t-t_0))}} \, .
\end{equation}
Again (\ref{P_KH}) correctly reproduces both the propagator for
the harmonic oscillator (for $\gamma=0$) and the propagator for
the free particle ($\gamma=0$ and $\omega=0$). However, for pure
damping both propagators are different.

There is a crucial difference between $K$ and $K_{\rm CK}$. The
Caldirola-Kanai propagator satisfies local composition law
\begin{eqnarray}\label{KKK-local}
  K_{\rm CK}(x_2,t_2;x_0,t_0) =\int dx_1 K_{\rm
    CK}(x_2,t_2;x_1,t_1)
    K_{\rm CK}(x_1,t_1;x_0,t_0) \ .
\end{eqnarray}
It is no longer true for $K$ defined in (\ref{PROPAGATOR}).
Therefore, as in the classical case memory term encoded into
dynamical equation is responsible for the breaking of the local
composition law for the quantum evolution.

Finally, let us consider a damped (without oscillations) evolution
of the initial Gaussian wave packet
\begin{equation}\label{}
   \psi_0(x;t_0) = C\, e^{-(x-x_0)^2/\sigma + ip_0 x/\hbar}\ .
\end{equation}
One easily finds for the probability density $\rho(x,t;t_0) =
|\psi(x,t;t_0)|^2$:
\begin{equation}\label{}
    \rho(x,t;t_0) = \frac{|C|^2}{\sqrt{\sigma(t-t_0)/\sigma}} \exp\left[ - 2\frac{(x - x(t))^2 }{ \sigma(t-t_0) }\right] \ ,
\end{equation}
where $x(t)$ is given by the classical formula
\begin{equation}\label{}
    x(t) = x_0 + p_0 \frac{\tanh(\gamma(t-t_0))}{m_0
    \gamma} \ ,
\end{equation}
and
\begin{equation}\label{}
    \sigma(t-t_0) =  \sigma \left[ 1 + \left(\frac{\hbar\tanh(\gamma(t-t_0))}{m_0 \sigma
    \gamma}\right)^2\right] \ .
\end{equation}
It shows that the presence of the damping term modifies the way
the initial way function spreads in time. Note, that in the
asymptotic regime one obtains
\begin{equation}\label{rho-a}
    \rho_{\rm asympt}(x;t_0) = \frac{|C|^2}{\sqrt{\sigma_{\rm asympt}/\sigma}}
    \exp\left[ - 2\frac{(x - x_{\rm asympt})^2 }{\sigma_{\rm asympt} }  \right] \ ,
\end{equation}
with the classical formula for the asymptotic position
\begin{equation}\label{}
    x_{\rm asympt} = x_0 + \frac{p_0}{m_0\gamma}\ ,
\end{equation}
and asymptotic dispersion
\begin{equation}\label{sigma-a}
    \sigma_{\rm asympt} = \sigma \left[ 1 + \Big(\frac{\hbar}{m_0\gamma\sigma}\Big)^2 \right] \ .
\end{equation}

 It is therefore clear that the damping
prevents the wave function from the total spreading. An initial
Gaussian probability density
 $\rho_0(x;t_0)$ cannot `spread more' than $\rho_{\rm
 asympt}(x;t_0)$. A similar phenomenon appears for Caldirola-Kanai
 dynamics  with different asymptotic state
\begin{equation}\label{}
    x^{\rm CK}_{\rm asympt} = x_0 + \frac{p_0}{2m_0\gamma}\ ,
\end{equation}
but with the same asymptotic dispersion $\sigma_{\rm asympt}$
given by (\ref{sigma-a}).

\section{Conclusions}

We performed an analysis of the   new equation of motion for the
damped oscillator (\ref{damped_NEW}). It differs from the standard
one by a damping term -- $2\gamma\tanh(\gamma(t-t_0))\dot{x}$ --
which is non-local in time and nonlinear in the damping constant
$\gamma$. The new parameter $t_0$ introduces effective memory. For
long time behavior $t \gg 1/\gamma$ one recovers standard equation
with a damping term -- $2\gamma \dot{x}$. Both classical and
quantum analysis is performed. The characteristic feature of this
nonlocal system is that it breaks local composition law for the
classical Hamiltonian dynamics and the corresponding quantum
propagator. Interestingly, the same propagator was recently
derived in \cite{Kochan}.  Without referring to Hamiltonian
formulation and using purely geometric methods  the author of
\cite{Kochan} derived the corresponding propagator starting from
the standard equation (\ref{damped}). We have shown that the
corresponding classical limit leads to the new damping equation
(\ref{damped_NEW}). Therefore, we conclude that the quantum
nonlocal propagator derived in \cite{Kochan} does originate in
nonlocal equation (\ref{damped_NEW}). Finally, it was shown that
the purely damped behavior modifies well known property of the
free quantum dynamics leading to the perfect spreading of the
initial wave function. It is no longer true when the dissipation is
present. Now, instead of the perfect spreading there is an
asymptotic state giving rise to the asymptotic probability density
(\ref{rho-a}).

\section*{Appendix}
\def\theequation{A.\arabic{equation}}
\setcounter{equation}{0}

Using a general formula for a propagator corresponding to
quadratic Hamiltonian (see e.g.~\cite{Kor}) one easily finds that
\begin{equation}\label{xp_Ham}
 H(x,p,t)\;=\;\frac{1}{2}\mu(t)p^2+\nu(t)xp+
\frac12 \lambda(t)x^2,
\end{equation}
with
\begin{equation}
 \mu(t)= \frac{1}{m_0\cosh^2(\gamma(t-t_0))}
\end{equation}
and
\begin{eqnarray}
 \nu(t) &=& \tanh(\gamma(t-t_0))\Big(\Omega\frac{\tanh(\gamma(t-t_0))}{\tan(\Omega (t-t_0))}-\gamma\Big)\,,\\
\lambda(t) &=& m_0\Omega^2\left\{ \frac{2\gamma}{\Omega}\,\frac{\tanh(\gamma(t-t_0))}{\tan(\Omega(t-t_0))}
+  \frac{1 - [1+\tanh^2(\gamma(t-t_0))]\cos^2(\Omega(t-t_0))}{
\sin^2(\Omega(t-t_0))} \right\}\ ,
\end{eqnarray}
produces (\ref{K-Kochan}), that is, it annihilates additional
phase factor $B(x,t;t_0)$ out of (\ref{PROPAGATOR}).

\end{document}